\documentclass[10pt,letterpaper]{article}
\usepackage[top=0.85in,left=2.75in,footskip=0.75in]{geometry}

\usepackage{amsmath,amssymb}

\usepackage{changepage}

\usepackage[utf8x]{inputenc}

\usepackage{textcomp,marvosym}

\usepackage{cite}

\usepackage{nameref,hyperref}

\usepackage{microtype}
\DisableLigatures[f]{encoding = *, family = * }

\usepackage[table]{xcolor}

\usepackage{url}
\usepackage{array}

\newcolumntype{+}{!{\vrule width 2pt}}

\newlength\savedwidth

\raggedright
\usepackage[aboveskip=1pt,labelfont=bf,labelsep=period,justification=raggedright,singlelinecheck=off]{caption}

\makeatletter
\renewcommand{\@biblabel}[1]{\quad#1.}
\makeatother

\usepackage{lastpage,fancyhdr,graphicx}
\usepackage{epstopdf}
\fancyheadoffset[L]{2.25in}
\fancyfootoffset[L]{2.25in}
\usepackage{hyperref}
\usepackage{comment}
\usepackage{graphicx}
\usepackage{makecell}
\usepackage{subcaption}
\usepackage[T1]{fontenc}
\usepackage{float}

\begin{document}
\vspace*{0.2in}

\begin{flushleft}
{\Large
\textbf\newline{A partial knowledge of friends of friends speeds social search}
}
\newline
% Insert author names, affiliations and corresponding author email (do not include titles, positions, or degrees).
\\
Amr Elsisy\textsuperscript{1,2,\Yinyang},
Boleslaw K. Szymanski\textsuperscript{1,2,\Yinyang,*},
Jasmine A. Plum\textsuperscript{1,2},
Miao Qi\textsuperscript{1,2},
Alex Pentland\textsuperscript{3},
\\
\bigskip
\textbf{1} Department of Computer Science, Rensselaer Polytechnic Institute, Troy, New York, United States of America, \textbf{2} Network Science and Technology Center, Rensselaer Polytechnic Institute, Troy, New York, United States of America, \textbf{3} Massachusetts Institute of Technology, Media Laboratory, Cambridge, Massachusetts, United States of America\\
\bigskip
\Yinyang These authors contributed equally to this work.\\
* szymab@rpi.edu

\end{flushleft}

\section*{Abstract}
Milgram empirically showed that people knowing only connections to their friends could locate any person in the U.S. in a few steps. Later research showed that social network topology enables a node aware of its full routing to find an arbitrary target in even fewer steps. Yet, the success of people in forwarding efficiently knowing only personal connections is still not fully explained. To study this problem, we emulate it on a real location-based social network, Gowalla. It provides explicit information about friends and temporal locations of each user useful for studies of human mobility. Here, we use it to conduct a massive computational experiment to establish new necessary and sufficient conditions for achieving social search efficiency. The results demonstrate that only the distribution of friendship edges and the partial knowledge of friends of friends are essential and sufficient for the efficiency of social search. Surprisingly, the efficiency of the search using the original distribution of friendship edges is not dependent on how the nodes are distributed into space. Moreover, the effect of using a limited knowledge that each node possesses about friends of its friends is strongly nonlinear. We show that gains of such use grow statistically significantly only when this knowledge is limited to a small fraction of friends of friends.

%linenumbers

\section*{Introduction}
Social search using Milgram rules has been extensively studied over the last 50 years. Briefly, the problem involves tasking a person to utilize direct social connections to send a folder to a target person. The next recipient should be a person who is most likely to increase the probability of the folder reaching the target. However, to avoid loops, any neighbor who previously held the folder is excluded from consideration. While there is a diverse body of work on this topic (see a thorough survey~\cite{schnettler2009structured}, the formulation used most frequently was defined in the initial Milgram's small world experiments~\cite{milgram1967small,travers1969experimental,korte1970acquaintance}). Milgram's work was notable for being the first empirical social experiment in which individuals used only contacts with whom they were on a first-name basis. We will refer to a social search using the above rules as {\it Milgram social search}. To make such search replicable, we emulate it on an actual social network, Gowalla in which information about the temporal locations of each user was available to researchers. Gowalla enables its users to spread information about location-based activities by {\it check ins} at physical locations that are broadcast to their Gowalla friends. Gowalla friends are connected by edges to which we refer as {\it friendship edges}. These edges define Gowalla communities, which we detect using a label propagation algorithm for community detection called LabelRank~\cite{xie2013labelrank}.
The goal is to conduct experiments that can be repeatedly run with changed parameters for forwarding rules and network topology to shed light on features of the search that are necessary to ensure efficient social search.

The distribution of the average shortest path lengths in online social networks has been measured. The majority of people in such networks are closer than six degrees of separation. Example distances are 4.7 in Facebook~\cite{backstrom2012four}, 3.7 in Myspace~\cite{ahn2007analysis}, 4.1 in Twitter~\cite{kwak2010twitter}, and so on~\cite{mislove2007measurement}. Such networks have been described as {\it small world} or {\it scale-free}~\cite{barabasi2000scale}. More formal definitions involve the average distance, diameter, the degree distribution, or their Power Law exponents, $\gamma$~\cite{barabasibook}. In these networks, there are {\it bridge} links that connect distant communities to each other. The idea of bridges, and furthermore that they are in some capacity {\it weak ties} has been suggested in literature~\cite{granovetter1973strength}, and is the basis for Milgram social search. 

The previous research \cite{dodds2003experimental} shows that each node selecting the friend to whom to forward the folder applies one of the three individual criteria. The first criterion is the candidate friend's distance to the target person. This is often the only criterion used for forwarding experiments that are inspired by Milgram's original experiment \cite{szule2014lost}. The second is this friend's membership in a community to which the target belongs. The third criterion is this friend's prominence (degree). Below, we describe a combined strategy that uses the value of a linear polynomial defined over the all three criteria metrics for selecting the best candidate friend for forwarding.  

An analysis of spatial distribution of friends reveals that in the Gowalla social network about 35\% of friends of an average user are located within a 160 km radius of that user's location~\cite{nguyen2013analyzing}. This fraction decreases for friends of friends (we will refer to them as {\it indirect friends} or {\it i-friends} in short), but still a significant 20\% of i-friends are within the same radius. However, for higher orders of indirect friendship, their fractions within the same distance drop below 5\%. 

Recent analyses of online social interactions confirmed the general conclusion about the spatial distributions of friends, but found that such distributions significantly differ inside and outside of metropolitan areas~\cite{laniado2018impact}. In general, people tend to interact with those that are geographically close, and as a result, the probability  of social interaction between peers decreases when the distance between them increases \cite{grabowicz2014entangling}. This motivated the author of reference~\cite{jordan2017link} to explicitly include this property in the network generator for spatially realistic social networks~\cite{ thurner2015physical}. 

Importance of indirect friends was observed in~\cite{phan2015natural}, in which two groups of students were subject of a study on interactions with i-friends. Hurricane Ike affected one group, but not the other. Affected students were more likely to connect with i-friends that were in close proximity than the members of the other group. This suggests that in times of need, users reach beyond their circle of friends to expand their knowledge base. This observation motivated us to check if use of nodes' awareness of i-friends can improve Milgram social search. However, to be realistic such use needs to avoid memory overload of each participant of a search. For this reason, a person is unlikely to know much about friends of each particular friend. So we cannot assume that that person would know the addresses of i-friends, or be on a first name basis with them. Thus, if partial knowledge of i-friends of a direct friend indicates that one of them is a good candidate to forward the folder, the sender can only pass the folder to that friend. After receiving the folder, that friend will be able to send it to the proper i-friend. Thus, reaching an indirect friend is a two-hop process. 

To account for the partial knowledge of i-friends, our model uses the parameter $\kappa$. It defines the maximum number of i-friends known to a person through each friend. If the number of friends that a friend has exceeds $\kappa$, then $\kappa$ known i-friends are chosen uniformly randomly. By simulations, we establish the optimal value of parameter $\kappa$ and coefficients for the linear polynomial with three criteria for the selection of the next node in social search. 

Finally, we pose two novel questions. The first is how much social search is affected by changes to $\kappa$ when forwarding decisions are based on partial knowledge of i-friends. We find that increasing $\kappa$ strongly and positively influences the performance of the social search only if $\kappa$ is small. The second novel question is how the different ways of distributing friendship edges, and use of partial knowledge of i-friends influence the Milgram social search. Surprisingly, we find that only the distribution of friendship edges affects efficiency of this search. Hence, preserving only this distribution and using partial knowledge of i-friends are necessary and sufficient conditions for efficient social search.

\section*{Results}

The crucial component of the design of our experiments is creating a realistic model of the current sender of the folder making selection of the successor. In reference \cite{dodds2003experimental}, the authors experimentally evaluate use of general criteria for these selections. We discuss their results in the \textit{Methods} section. Here, we describe how we use these criteria in our model for successor's selection. 

Whenever a user receives a folder to forward, it computes the following utility score $U_i$ for every friend $i$ (for brevity, we do not explicitly list argument $i$, when it is clear from the context):
\begin{equation}
\label{eq:score}
U_i = W_D*D_m + W_C*C_m + W_P*P_m,
\end{equation}
where $W_D$, $W_C$, and $W_P$ are rational weights in the range $[0,1]$. Except for random search in which all weights are 0, $W_P=1-W_C-W_D$. 
Normalized distance metric, $D_m$, denotes the normalized distance between the locations of node $i$ and the target. Community metric, $C_m$, defines the normalized size of the community to which both the node $i$ and the target belong. Prominence metric, $P_m$, denotes the normalized degree of node $i$. These metrics are defined in the \textit{Methods} section. Then, the user holding the folder sends it to the friend with the highest score, chosen randomly if there is more than one node with such score. 

Using categories of metrics used by users discussed in \cite{dodds2003experimental}, we define a selection metric based on these categories.
Let $D_{max}$ stand for the diameter of the network, and $D_i$ denote the distance from node $i$ to the target. 
If there is a community to which both the target and the current holder of the folder belong, we set $C_{max}$ to the size of this community, otherwise $C_{max}$ is set to $N$, the size of Gowalla network. In our study, Gowalla communities were detected using a label propagation algorithm named Label rank \cite{xie2013labelrank}.  
Finally, let $P_{max}$ denote the largest node degree in the network, and $P_i$ be the degree of node $i$. Then, the metric values for node $i$ in Eq.~\ref{eq:score} are defined as follows:
\begin{eqnarray}
\label{eq:partials}
D_m & = & 1 - \frac{\log(D_i)}{\log(D_{max})}\\
C_m & = & 1 - \frac{\log(C_{max})}{\log(N)}\\
P_m & = & \frac{1}{\log(P_{max})-\log(P_i)+1}
%better P_m & = & \frac{\log(P_i)}{\log(P_{max}}
\end{eqnarray}

For each experiment run on one of the five configurations, we record its success rate of delivering the folder to the target user. Fig.~\ref{fig:Multiper} shows the success rate as a function of the maximum number of i-friends allowed to be known to each node for each of its friends ($\kappa \in [0,48]$). The thick color plots show boundaries of one standard deviation from the average taken over 500 runs for each of the five configurations used in the experiments. The first configuration, with all weights equal to zero, randomly chooses a friend to whom to forward a folder and serves as a baseline. However when $\kappa>0$, even in this case the forwarding node checks if any of its known i-friends is the target node. If it is, the search ends successfully by forwarding the folder to the target through the mutual friend of the current folder holder. This, as seen in the plot, increases the success rate quite significantly compared to pure random forwarding.
The next three configurations are those with the boundary values of weights, which are 1.0 for one metric, and 0.0 for the remaining two metrics. The fifth configuration of $W_D=1/12, W_C=7/24, W_P=5/8$ performs the best in the original search. We found the optimal weights by conducting a binary search starting with $W_D = W_C = W_P = 1/3$, and then varying each weight initially by $\Delta$ of $1/3$, and then by halving $\Delta$ at each step, while maintaining the sum of all weights at $1$. We use these five configurations to measure the impact of $\kappa$ on the behavior of the system over a broad range of search criteria. Starting with the search without i-friends awareness ($\kappa=0$), as $\kappa$ increases, delivery rate improves, but this trend becomes less pronounced for larger $\kappa$. With the optimal weights $W_D=1/12, W_C=7/24, W_P=5/8$ and $\kappa= 15$ the success rate of search is 94\%. However, increasing $\kappa$ over 15 barely improves the success rate. It should be noted that the average number of friends per user is 11.98 friends. Of the total of 75,803 users, 48,844 of them have less than 15 friends but more than one friend, so more than 70\% users may benefit from having knowledge of their i-friends. The average number of distinct i-friends per user is
2,016, but with the limit of $\kappa=15$ imposed, this number drops by order of magnitude to 125.3.
\begin{figure}[htb]
\centering
\caption{Success rates with error bars collected from 500 runs for each of the five selections of search weights defined in Eq.~\ref{eq:score} with different seeds for the random number generator for each of 500 distinct pairs of starting and target  nodes selected to be at least 1,609 km apart.
The plots include the baseline random search with all metric weights set to 0, three searches using a single metric with weight 1, and the search with the metric weights yielding the highest performance. We plot each rate as a function of the maximum number of i-friends of which a node is aware for each of its friends. The error bars show the standard error.}
\label{fig:Multiper}
\end{figure}
Another significant result is finding the importance of distribution of friendship edges. This agrees with a finding from \cite{Liben2005geographic} that the efficient Milgram social search requires geographic based friendship edge distribution.
Distributing friendship edges involves two steps. First, each node randomly but according to the selected distribution, chooses its degrees $d$ (which means how many friends this node will have) according to an appropriate distribution. Then, a node selects $s$ times a random friend to connect to by a friendship edge.

To efficiently deal with space distribution of nodes and their friends, we cover the contiguous U.S. territory (i.e., excluding Hawaii and Alaska) by an array of non-overlapping, approximately equal-size rhomboids with 70 km sides (for simplicity, we refer to them as squares below). We draw rhomboid sides along meridians and parallels to simplify translation of geographical coordinates into positions in a rhomboid and vice versa. Initially, we cover the U.S. territory with 1,860 rhomboids, many of which have no Gowalla users inside them. We remove rhomboids without Gowalla users, and only process in experiments the remaining 850 rhomboids.

When studying empirical complex networks with structures and properties that evolve naturally, it is often challenging to understand which elements of a structure are essential to the observed dynamics and properties. To address this challenge, an approach was developed of shuffling nodes, edges, and time sequences of dynamic events, and then simulating resulting process dynamics to test if the properties or features of interest are preserved. For example, the authors of 
reference~\cite{barabasi1999emergence} study what properties are necessary for evolution of a network to end in a scale-free network. The authors demonstrated that both growth and preferential attachment are necessary for such evolution. In reference~\cite{sinatra2016quantifying}, the authors attempt to predict the timings of scientists publishing their most cited paper from the sequence of the author's all publications. The question arose if there are detectable changes in citation to the scientist's papers at time leading up to, or following the publication of the highest cited paper. To test it, the authors shuffled the sequences of publications of all scientists. The shuffled sequences were similar to the original ones, answering the question negatively. The authors of reference \cite{jia2017quantifying} study how scientists change the focus of their research over time. The authors performed three shuffling experiments to test three properties of sequences of publications. For brevity, we described a test of recency. Shuffling the order of publication sequences resulted in disappearance of recency. This proved that selecting next topic researchers are more likely to return the most recent ones rather than older ones. This is a property that shuffling easily destroys. Following these lines of experimenting, we use shuffling connectivity and geographical locations of nodes in Gowalla network to test which distributions of friendship to edges and nodes into space preserve or disturb efficiency of social search.

To quantify how the distribution of Gowalla users over space influences social search, we use eight distributions of nodes into space. The original Gowalla distribution constitutes the baseline. The next method is random distribution, which assigns a location to each Gowalla user by first selecting a rhomboid uniformly randomly, and then choosing two orthogonal coordinates within it for placement of the user. We generated 10 samples with this distribution.
The last six distributions combine one of the three distributions of nodes to rhomboids with one of two methods for embedding individual nodes into the space of each rhomboid. The first three are respectively exponential, normal and Zipf distributions, each with the mean of the rhomboid population in the original data.
The normal distribution uses the variance of rhomboid population in the original Gowalla network. The Zipf distribution starts with the largest Gowalla user population in a single rhomboid of $10,700$, which yields the closest total population to the total number of Gowalla users. Again, we generated 10 samples for each distribution.
The first individual node embedding into space is geographic, which places users in the given rhomboid using positions occupied by real users in one of the original rhomboids whose population is close to the given rhomboid population. The second one embeds the individual users uniformly randomly into the rhomboid space. We generated 10 samples of each distribution resulting in 100 samples for each considered distribution.

To measure the influence of choice of distribution of friendship edges on the results, we use the following five methods for assigning node degrees to Gowalla users.
The first is the original friendship edge distribution used as a benchmark. The second uses the random distribution preserving degree/range of friends while randomizing friendships. To achieve that, each node swaps edges with its neighbors in the same rhomboid with friends in the same range of distances to those nodes. The random uniform friendship edge distribution generates an Erdős–Rényi random graph, which has the same node average degree as the original Gowalla network does. We generate 10 samples of this distribution. The last two distributions assign each node a degree according to the exponential and Power Law distributions with the mean degree of the original graph. In addition, the Power Law uses the node degree exponent $\gamma=1.49$ of the original Gowalla graph, and the range of node degrees selected to closely match the total number of generated nodes with  the number of nodes in the original Gowalla network. Then, we use the created degree sequences to generate sample friendship graphs with these distributions. We generated 10 samples of each distribution resulting in 100 samples for each case of the considered friendship edge distributions.
We ran each created sample 10 times and averaged the results. Then, we implemented and ran the elch's \cite{welch1947generalization}, two-tailed, t-test to check whether or not the differences in performance are statistically significant, and we report the results below.

\subsection*{Success rate and awareness of i-friends}

Even with limited awareness of i-friends, the spatial distribution of nodes did not really matter for the success rate of the Milgram social search Fig.~\ref{fig:SSFoF}(a). In contrast, the distribution of friends has a major impact on success rate. The original set of friendships achieved the highest success rate that, surprisingly, is independent of the way nodes are distributed over the space. The success rate with partial awareness of i-friends for $\kappa=15$ is statistically significantly higher than the success rate with $\kappa=0$ (no knowledge of i-friends) and all other friendship edge distributions (P-value$=0.0005$). The distant second is the random friendship edge distribution preserving the degree and the ranges of distances from friends of each node, whose success rate is not statistically significantly higher for this distribution combined with $\kappa=0$ (P-value$=0.2481$), but it is statistically significantly higher  for the remaining tested methods of friendship edges distributions (P-value$\leq 0.0005$). The remaining three methods of distributing friendship edges achieve much lower success rates. Accordingly, their success rates without knowledge of i-friends are even lower and since they drop below 10\%, they are not shown. 

\begin{figure}[h!]
\centering

\caption{\textbf{(a-b)} show plots with error bars of success rates (a) and stretches (b) achieved with partial knowledge of i-friends under the different distributions of friendship edges as a function of various distributions of nodes into space. Plots represent the results of running 10 samples of each distribution resulting in 100 samples for each case of the considered friendship edge distributions. Each of these samples was executed 10 times and averaged results plotted.
Each friendship edge distribution has a unique color assigned to its plots and two best performing distributions have also plots of their stretches achieved without knowledge of i-friends marked with the dashed line. We describe all distributions of friendship to edges and nodes into space in the text. Plots for runs with awareness of i-friends were computed using $kappa$ limit of the number of i-friends set to 15, which, if needed, are uniformly randomly chosen from i-friends for each friend of the sender. The error bars were in the range of [0.002, 0.039] for success rates (a) and in the range of [0.07, 1.61] for stretches (b).}
\label{fig:SSFoF}
\end{figure}

The overall conclusion is that the distribution of friendship edges and use of i-friends are important, while the spatial distributions of nodes are not. The success rate achieved on the original Gowalla network with the i-friends awareness was 94\%, so higher than the 77.8\% rate achieved without it. The difference is significant from the perspective of the failure rate, which is 6\% in the first case but 22.2\%, so over three times higher than in the first case. 

\subsection*{Path stretch and levels of i-friends' awareness}

The stretch of the shortest distance between nodes $n_1, n_2$ is defined as $s_{n_1,n_2} = {\bar d}(n_1, n_2)/d_s(n_1,n_2)$, where ${\bar d}$ is the average distance traveled and $d_s$ is the length of the shortest path. As explained below, if a transfer to an i-friend is used, it adds two steps to $d_s$ because two hops are required to pass the folder from the current folder to the i-friend. Thus, the closer the stretch is to $1$ the more efficient the search is. As Fig.~\ref{fig:SSFoF}(b) shows, stretch is the lowest for the original set of friends. The distant second is the random friendship edge distribution preserving for each node the degree and the ranges of distances from friends. For the remaining three types of random friendship edge distributions, exponential, uniform and Power Law, the stretch drastically increases. 

The lowest stretch achieved with the i-friends awareness is 1.82 with original friendship edge distribution, lower than the 2.72 stretch observed in experiments without such awareness. In the first case, the stretch includes about 94\% of all paths while in the second 77.8\% paths, since there are 16.2\% more paths on which the second case failed. To fairly compare the stretches between these two cases, we compute a stretch with the i-friends awareness only in cases in which search without i-friends awareness succeeded. This adjustment drops the stretch to 1.59, showing the great improvement over forwarding without the i-friends awareness.

In conclusion, when i-friends awareness is used, the distribution of nodes does not affect the stretch, but the friendship edge assignment does.

Like in Fig.~\ref{fig:SSFoF}(a) we also show the stretch with and without i-friends awareness for the second best performing distribution of friendship edges that again is the original and the random preserving nodes' degree and ranges of distance to their friends. The stretch of the original friendship edge distribution with $\kappa=15$ is statistically significantly lower then that of this distribution with $kappa=0$ (P-value$=0.0051$), and the remaining friendship distributions (P-value$\leq 0.0005$). Similarly, the stretch of the second in performance random friendship edge distribution preserving the degree and the ranges of distances from friends of each node is not statistically significantly higher for this distribution combined with $\kappa=0$ (P-value$=0.5038$), but it is statistically significantly higher  for the remaining tested methods of friendship edges distributions (P-value$\leq 0.0053$).

\section*{Methods} \label{sec:Methods}
 The data for building Gowalla social network used here was originally collected by some of the authors using publicly accessible Gowalla's API for a study of the location-based social networks~\cite{nguyen2012using}. All collected data were anonymized according to the protocol approved by the RPI Institutional Review Board (IRB). At the time of collection, Gowalla was a global network with users primarily located in the United States and Sweden, and contained 154,557 users (nodes) and 1,139,110 friendship edges distributed according to the Power Law with node degree exponent $\gamma\approx1.51$. 

Since there are large differences between numbers of Gowalla users and populations in countries outside of the United States, here we analyze only data for users located in the U.S. To ensure connectivity, we only consider the giant component of the analyzed network that comprises 75,803 users and 454,350 friendship edges, with the node Power Law degree exponent $\gamma\approx1.49$.  In Fig.~\ref{fig:PowLawDistGowalla}, we plot the degree distribution for the giant component of a network comprised of Gowalla users located in the U.S.

\begin{figure}
    \centering
%    \includegraphics{degree_distribution.png}
%Fig. 3
    \caption{The degree distribution for the giant component of a network with 75,803 Gowalla users located in the U.S., 454,350 friendship edges, and $\gamma\approx1.49$.}
    \label{fig:PowLawDistGowalla}
\end{figure}

For each run of social search emulation, we randomly uniformly select the starting user and the target user, which are at least 1,609 km apart, and then execute a Milgram social search for up to 50 hops. In each hop, the user currently sending the folder selects one friend as the next recipient.
The search ends successfully when the target receives the folder.

As mentioned above, in reference \cite{dodds2003experimental}, the authors evaluate several criteria for selecting the node to whom to forward the folder. This reference identified nine such criteria, of which we skipped two: {\it others}, which mainly includes the target nodes and {\it continue the chain}, which is difficult to categorize. We group the remaining seven criteria into three categories of nodes: (i) nodes that are the closest distances from the target, (ii) nodes with the highest degrees, and (iii) nodes belonging to the community to which the target node also belongs. The first category includes {\it geography} used by 35\% of chains, {\it traveled to the target's location} employed by 14\% of chains, and {\it family originated from the target's location} used by 11\% of chains. Hence, 60\% of chains used category distance of which 25\% used i-friends' awareness. Just 8\% of chains used category prominence that includes only one criterion: {\it lots of friends}. Finally, 9\% of chains used {\it work}, 8\% of chains used {\it similar profession} and 4\% of chains used {similar education}, which are all members of the third category. Thus, 21\% of chains used this category. Interestingly, 56\% of successful chains used the distance during the first four steps of search. Later, its use in steps 5-7 dropped to 29\%. In contrast, 57\% of successful chains used community sharing in steps 5-7. Earlier, its use in steps 1-4 was just 28\%. The prominence usage was low in successful chains, with 7\% usage in early steps and 6\% in later steps. The authors did not discuss reasons for low usage of prominence in their experiments.  The possible culprit might be limited knowledge of the degrees of the neighbors in the studied network. Email networks typically do not make the sender aware of the recipient's degree. Another interesting point is that criteria {\it traveled to the target's location} and {\it family originated from the target's location} actually select friends based on awareness that these nodes are likely to have some friends in the target's location. The use of such indirect awareness of this type of i-friends in the real experiments shows that some participants of this experiment intuitively recognized the value of awareness of i-friends. Thus, some people engaged in real Milgram social search have already used the i-friends awareness as postulated here. Our contribution is to elevate such awareness to an explicit criterion of choice comparable to distance, prominence or shared community and to demonstrate the improvement that such awareness yields. This approach has been used to improve design of routing protocols for delay 
networks ~\cite{sagduyu2017multilayer} and most recently for IOT~\cite{yang2020social}.

All values $\kappa$ from 0 to 48 with step 3, and all combinations of $(W_D, W_C, W_P) \in \{(0,0,0),(1,0,0), (0,1,0),(0,0,1),(1/12,7/24,15/24)\}$ are used in a set of computational experiments. For each of them, we select uniformly randomly 500 pairs of starting and target nodes under the condition that they are at least 1,609 km apart. 
We average the results for each pair of users over 500 runs. Selecting i-friends, we use the principle of coordinated execution~\cite{jankowski2018probing}, which ensures that for a given set of weights $W_D, W_C, W_P$, every node selects the same i-friends in each experiment repetition. The coordinated execution was also used for experiments in which the same weights were used but with different $\kappa$ values. As a result, each set of i-friends selected with large $\kappa$ value contains all i-friends selected with smaller $\kappa$ values. In short, for the given graph, let $F(\kappa)$ denote a set of i-friends selected with $\kappa$, and let $\kappa_1\leq\kappa_2$ hold, then $F(\kappa_1)\subseteq F(\kappa_2)$. This ensures fair comparison of results with different values of $\kappa$.

\section*{Discussion}

Since U.S. metropolitan areas are populated by a large fraction of the U.S. total population, large concentrations of 
Gowalla users also reside there. Table~\ref{tab:GowallaFractionalPopulation} shows strong correlations between populations of the Gowalla users  and the inhabitants of the U.S. metropolitan areas.
Yet, there are some differences. There are up to 11\% more Gowalla users in several metropolitan areas (e.g., Austin and San Antonio, San Francisco and San Jose) than their share of the U.S. population.
However, there are a few areas underrepresented that way with a small deficit, reaching at most about 2\% of the population. 

\begin{table}[h!]
\centering
\begin{tabular}{|c|l|r|r|r|}
\hline
No. & Name & \thead{Percentage of \\ U.S. population} & \thead{Percentage of \\ Gowalla Users} & \thead{Difference}  \\
\hline
1 & Baltimore-Washington DC & 2.46 & 2.93 & -0.47 \\
\hline
2 & Los Angeles & 7.42 & 1.22 & 6.21 \\
\hline
3 & Dallas-Fort Worth & 6.97 & 2.18 & 4.79 \\
\hline
4 & Austin and San Antonio & 12.71 & 0.97 & 11.74 \\
\hline
5 & Seattle-Tacoma-Belly & 2.06 & 1.17 & 0.89 \\
\hline
6 & New York City & 4.25 & 6.30 & -2.05 \\
\hline
7 & Boston & 1.55 & 1.48 & 0.06 \\
\hline
8 & Houston & 2.04 & 2.04 & 0.01 \\
\hline
9 & San Francisco and San Jose & 7.75 & 1.44 & 6.31 \\
\hline
10 & Chicago & 1.89 & 3.00 & -1.12 \\
\hline
11 & Philadelphia & 1.01 & 1.90 & -0.89 \\
\hline
12 & Salt Lake City & 1.14 & 0.36 & 0.78 \\
\hline
13 & Portland & 1.19 & 0.74 & 0.46 \\
\hline
14 & Denver & 1.35 & 0.88 & 0.47 \\
\hline
15 & Atlanta & 1.60 & 1.98 & -0.37 \\
\hline
16 & Oklahoma City & 1.82 & 0.41 & 1.40 \\
\hline
17 & Orlando & 2.77 & 0.73 & 2.04 \\
\hline
\end{tabular}
\caption{\label{tab:GowallaFractionalPopulation}Percentages of Gowalla users and the populations of metropolitan areas in the United States. 
}
\end{table}

This conclusion is based on analysis of Table~\ref{tab:FriendDistanceDistrUSO} that shows the fraction of friendships and indirect friendships (i-friends) whose geographical separation is within a given distance range, as well as the fraction of communities within a given average distance range of their members. The data demonstrates that distributions of communities, friends and i-friends over space are unique, providing complementary ways for reaching targets.

\begin{table}[h!]
\centering
\begin{tabular}{| c | r | r | r | r | r | r |}
\hline
 Range &  \multicolumn{3}{c|}{Percentage of}  & \multicolumn{3}{c|}{Cumulative Percentage of} \\
\hline
(km)        &  Friends  & i-\ \ \ \ \  & Commu-  &  Friends  & i-\ \ \ \ \  & Commu- \\
            &            & friends    & nites   &           & friends    & nities \\
\hline
$\leq$ 6.25 & 18.6      & 2.6 & 14.0    & {\bf 18.6}& 2.6 & 14.0 \\
\hline
6.25 -- 12.5 & 8.6     &  1.3 & 4.3    & {\bf 27.2} &3.9 & 18.3 \\
\hline
12.5 -- 25 & 10.3   & 1.7  &5.5     & {\bf 37.6} & 5.6 & 23.9 \\
\hline
25 -- 50 & 7.6    & 1.5  &4.6     & {\bf 45.2} & 7.1 & 28.5 \\
\hline
50 -- 100 & 3.9   & 1.0  & 2.6    & {\bf 49.0} & 8.1 & 31.1 \\
\hline
100 -- 200 & 3.8  & 1.6  & 3.1    & {\bf 52.8} & 9.7 & 34.1 \\
\hline
200 -- 400 & 6.4  & 6.0  & 6.8    &      59.2 & 15.7 &{\bf 40.9}\\
\hline
400 -- 800 & 6.4  & 8.8  & 8.2    &      65.6 & 24.5 &{\bf 49.1} \\
\hline
800 -- 1600 & 11.8&23.8  & 17.2   &      77.4 & 48.3 &{\bf 66.4} \\
\hline
1600 -- 3200 &14.8&36.4  & 23.3   &      92.2 & {\bf 84.7} &{\bf 89.7} \\
\hline
3200 -- 6400 &7.5 &14.2  & 10.0   &      99.8 & {\bf 98.9} & 99.7 \\
\hline
\end{tabular}
\caption{\label{tab:FriendDistanceDistrUSO} Distributions of fractions of friends, i-friends and communities over the ranges of distances from nodes. 
The second column shows the fractions of friends at each distance range, computed by summing the numbers of friends in each range for each individual user and then dividing the result by the total number of friends of all users.  
The third column shows fractions computed as ratios of sums of the numbers of i-friends of each user at each distance range to the total number of i-friends for all users. The fourth column shows fractions of members of communities of each individual user at each distance range listed, computed as the sum of these numbers divided by the total number of members of all relevant communities. Fifth, sixth and seventh columns list cumulative values from the second, third and fourth column, respectively.}
\end{table}

Over half ($52.8\%$) of friendships are within 200 km range from a user, the same distance constraint would not cover many of the communities in the network (only $34.1\%$ of them are covered) or i-friends (just $9.7\%$ of those are covered). On the other hand, over half ($50.6\%$) of i-friends are within the far range, from 1,600 to 6,400 km, while for communities, over half of their members ($55.5\%$ exactly) are in the middle range from 200 km to 3,200 km. These statistics show that the three metrics for making decisions at a given node have complementary information about the nodes at different ranges of distances from that node. For example, the main difference between spatial distribution of friends and i-friends is that only about $47.2\%$ of friends are located at a distance of 200 km or more while nearly twice as large percentage, 90.3\% to be exact, of i-friends are there.

It is important to consider how much information people can retain about friends of their friends, and how human memory limitations affect social search. The first question arises because of the sheer size of the sets of i-friends. The average Gowalla user has 12 friends, but 2,016 distinct i-friends, which is due to prominent users, with as many as 8,000 friends, being more accessible through the use of i-friends. To address this concern, for each friend we limit to 15 the number of i-friends which are, if needed,  uniformly randomly chosen from friends of this friend. This reduces the average number of friends of all friends of which a person is aware by order of magnitude, to 125.3, still not a small number. Fortunately, each of the metrics used in social search needs only very limited knowledge about any chosen friend of a friend. For distance, it is sufficient for a person to know something about travel destinations or past residences of each friend, especially if those are distant or unusual places. Likewise, for communities, it is reasonable to assume that a person knows the friend's attributes such as profession, interests, hobbies, and, thus, can associate some friends of this friend with an appropriate community. In case of prominence, it is likely that a friend occasionally mentions some notable or prominent friends, the existence of which the listener will remember more vividly than when ordinary friends are mentioned.

Such limited awareness of i-friends allows the node with a folder to send it to a friend of whose friends the sender is aware have a high chance to be the best choice for reaching the target. We estimate that the amount of such knowledge about a friend of that friend is at most $5\%$ of the amount of information about the corresponding friend. With the average of 15 i-friends per friend, the amount of such information about all i-friends is less than the amount of information held about each direct friend.

To demonstrate how the awareness of i-friends improves search, we analyze its impact on search metrics. We start with the distance. Let the distance from the node $n$ holding the folder to the target, $t$, be $r$ km. An annulus defined by two circles with radii of $3r/2$ and $r/2$, centered at node $n$ has the area of $2\pi r^2$. The circle centered at node $t$ of radius $r/2$ and area $\pi r^2/4$ contains all nodes that are distant at most $r/2$ from the target and it contains on average 1/8 of all nodes in the considered annulus. A randomly chosen point in this circle has an average distance to the target of $r/3$. With the average number of friends of a node being 12, the cumulative fraction of friends counted from the most distant annulus to the closer ones to the current folder sender is at least 2/3. For the annulus with an outer distance of 25 km, the expect number  of noode inside the circle around the target is at least according to Table~\ref{tab:FriendDistanceDistrUSO}. Thus, The expected reduction of the distance to target in a single step is 500 km.

The average distance between the starting and target node is 2,642 km. At this distance, the target is between two last circles, the inner circle with a radius of 1,600 km and the outer circle with a radius of 3,200 km. The corresponding annulus contains $36.4\%$ of all i-friends. With an average number of $125.3$ i-friends for the node currently holding the folder, the expected number of i-friends is $5.7$ in the circle of 940 km from the target. Thus, the expected distance of the closest i-friend to the target in this circle is 210 km. It takes on average more than three steps to get so close to the target using direct friends' knowledge, while using i-friends, one operation with the cost of two steps suffices.
Hence, a fraction of i-friends needed to have at least one i-friend inside the circle around the target has to exceed $8/125.3=6.4\%$, which is smaller than the fraction that the most distant annulus contains. 
Thus, the average reduction of the distance to target in this case is at least 3,200 km, achieved in two steps; the first step sends the folder to a friend and the second directs the folder to the appropriate i-friend. Hence, the gain of the distance towards target using partial i-friends awareness is over three times larger than using knowledge of direct friends.

For the community metric, the average number of communities to which a node belongs is nearly for two reasons. First, the algorithm that we used to find Gowalla communities assigns each node to at most one community. Second, the majority of nodes are members of a community. In Gowalla, the average number of communities that can be reached via direct friends is $6.8$ in a step and $13.6$ in two. For communities reachable using i-friends of which nodes are aware, this number grows to $47.4$ in two steps, so about 3.5 times more communities than reachable just by friends.

In the case of prominence, we distinguish between prominent nodes, which are those whose node degrees rank at the top $1\%$ of all nodes, and the non-prominent nodes that do not satisfy this condition. There are $758$ prominent nodes in Gowalla network, each with a degree of at least $122$. We naturally exclude those nodes from analysis as they have many friends so they are unlikely to use i-friends' prominence for search. So if all direct friends of the current folder sender are non-prominent, we established that they will have on average $19.3$ direct friends of which $4.4$ will be prominent. However in such a case, using i-friends of which the sender is aware, this number grows to $75.2$ i-friends of which $15.4$ are prominent, which amounts to a 3.5 times increase in knowledge of prominent nodes.

Looking at the results of the social search simulations run on the original Gowalla network, we can make some interesting observations. 
The results of social search conducted in the original Gowalla network, 60\% of the hops used direct friends and significant 40\% used i-friends. It is also important to note that even though we allow for up to 15 i-friends for each friend, an average of only 3.5 i-friends are actually used. We will call a metric dominating in a forwarding decision, if it is the largest component of the total score of the selected node. When using direct friends, the distance metric dominated in 0.3\% of hops, the community metric in 22.7\% of hops, and the prominence metric in 77.0\% of hops. When using i-friends, the distance metric dominated 0.6\% of hops, the community metric 30.1\% of hops, and the prominence metric 69.3\% of hops. 

When a sender directs the folder to a friend intended for its friend, the recipient follows this intention in 74\% of the cases. In 22\% of the cases, the recipient sends the folder to a friend who has a friend best fitting to get the folder to target. In the remaining ~4\% of the cases, the folder ends up at the direct friend of the recipient different from the intended one.

\section*{Conclusions} \label{sec:conclusions}
We make two contributions to understanding Milgram social search efficiency. First, we strengthen the result presented in reference \cite{dodds2003experimental} that geographical friendship edge distribution is {\it sufficient} to make the Milgram social search efficient. Using massive computational experiments, we demonstrate that the distribution of friendship edges and use of partial knowledge of i-friends are both {\it necessary and sufficient conditions} for social search efficiency.  

The second contribution is a discovery that awareness of the sender's i-friends is very beneficial for social search. It extends the information base of the sender about connections beyond the direct links to the sender's friends. It also improves the user's ability to identify friends whose friends have information independent from one held by the sender's direct friends. Furthermore, increasing such awareness when it is small brings significant improvement to social search efficiency.  This approach has been used to improve design of routing protocols for delay networks~\cite{sagduyu2017multilayer} and most recently for IOT~\cite{yang2020social}.

%\nolinenumbers

\ \ \ 

\noindent
{\bf Data Availability Statement:} All relevant data collected from Gowalla used in the paper are available at github repository at 
\url{https://github.com/amrelsisy/Gowalla-Data}.

\ \ \ 

\noindent
{\bf Acknowledgment:} The authors were partially supported by: the Army Research Laboratory under Cooperative Agreement Number
W911NF-09-2-0053 (the ARL Network Science CTA), the Army Research Office, Grant W911NF-16-1-0524, and the Office of Naval Research, Contract N00014-15-1-2641. The funders had no role in study design, data collection and analysis, decision to publish, or preparation of
the manuscript.

\ \ \ 

\noindent
{\bf Competing interests:} The authors have declared that no competing interests exist.

%\bibliography{main}

\end{document}